%% file: 0_main.tex
\documentclass[manuscript,screen]{acmart}
\usepackage{tabularx}
\usepackage{verbatim}
\usepackage{comment}
\usepackage{xcolor}

\acmJournal{PACMHCI} 
\acmYear{2024}       
\acmVolume{1}        
\acmNumber{CSCW2}    
\acmArticle{XX}      
\acmMonth{10}        

\acmConference[CSCW '25]{The ACM Conference on Computer-Supported Cooperative Work and Social Computing}{October 18--22, 2025}{Bergen, Norway}
\acmBooktitle{Proceedings of CSCW '25: The ACM Conference on Computer-Supported Cooperative Work and Social Computing}

\AtBeginDocument{%
  }

\setcopyright{acmlicensed}
\copyrightyear{2024}
\acmYear{2025}
\acmDOI{XXXXXXX.XXXXXXX}

\acmBooktitle{Proceedings of ACM Transactions on Accessible Computing}
\acmISBN{978-1-4503-XXXX-X/25/10}

\begin{document}

\title{AI-Mediated Hiring and the Job Search of Blind and Low-Vision Individuals}


\author{Kashif Imteyaz}
\email{imteyaz.k@northeastern.edu}

\affiliation{%
  \institution{Northeastern University}
  \city{Boston}
  \state{Massachusetts}
  \country{USA}
}

\author{Qiushi(Anya) Liang}
\email{liang.qiu@northeastern.edu}

\affiliation{%
  \institution{Northeastern University}
  \city{Boston}
  \state{Massachusetts}
  \country{USA}
}

\author{Yakov Bart}
\affiliation{
 \institution{Northeastern University}
 \city{Boston}
 \state{Massachusetts}
 \country{USA}}
 
 \author{Maitraye Das}
\affiliation{%
  \institution{Northeastern University}
  \city{Boston}
  \state{Massachusetts}
  \country{USA}}
\email{ma.das@northeastern.edu}

\author{Saiph Savage}
\affiliation{%
  \institution{Northeastern University}
  \city{Boston}
  \state{Massachusetts}
  \country{USA}}
\email{s.savage@northeastern.edu}


\renewcommand{\shortauthors}{Imteyaz et al.}

\begin{teaserfigure}
  \centering
  \includegraphics[width=\textwidth]{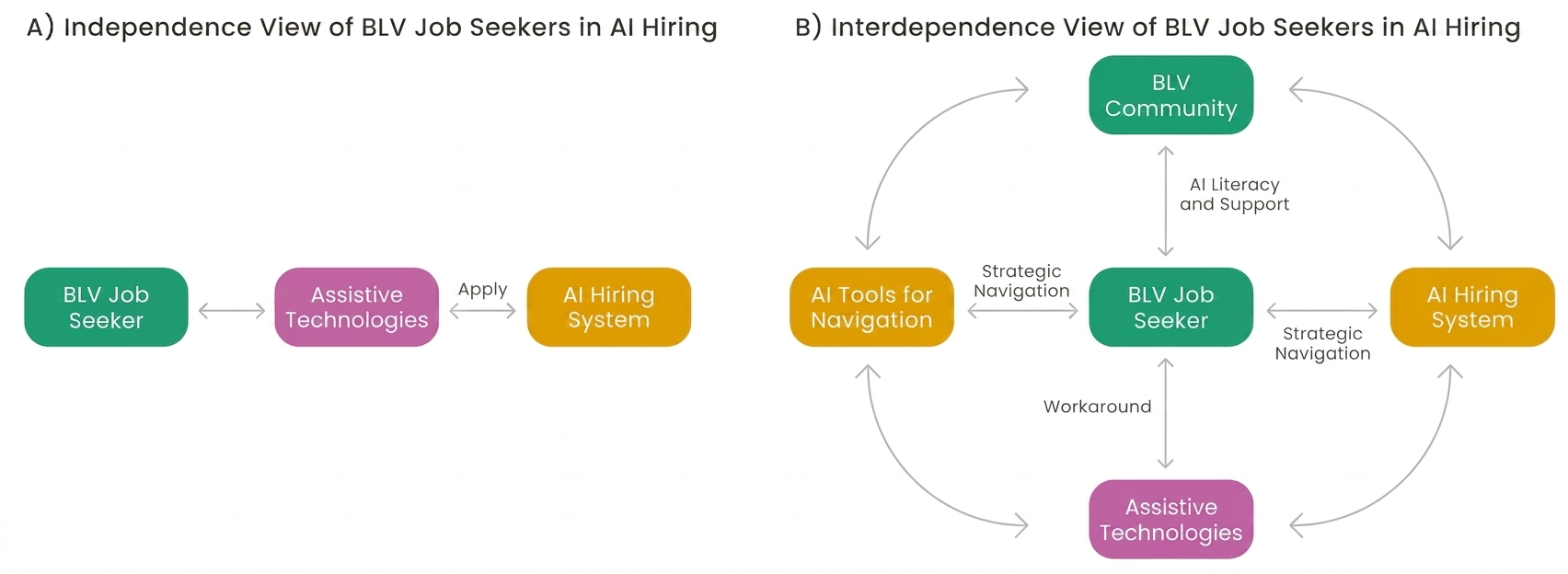}
  \caption{Conceptual models of BLV job search: (A) Independence and (B) Interdependence views.}
  \label{fig:comm034}
\end{teaserfigure}
\begin{abstract}
 Blind and low-vision (BLV) individuals face high unemployment rates. The job search is becoming harder as more employers use AI-driven systems to screen resumes before a human ever sees them. Such AI systems could inadvertently further disadvantage BLV job seekers, introducing additional barriers to an already difficult process. We lack understanding of BLV job seekers' experiences in today's AI-driven hiring ecosystem. Without such understanding, we risk designing technologies that create new systemic barriers for BLV job seekers rather than providing support. To this end, we conducted interviews with 17 BLV job seekers and analyzed their experiences with AI-powered hiring systems.  We found that AI hiring systems misrepresented their professional identities and created dehumanizing interactions. To level the playing field, BLV job seekers used strategic counter-navigation: they deployed their own tools to bypass algorithmic screening and built peer networks to share AI literacy. They also practiced ``strategic refusal,” choosing to avoid certain AI systems to regain their agency. Unlike prior work that frames job search as an individualistic activity, or one focused on being compliant with employer needs, we use the interdependence framework to argue that for BLV people, job search is an interdependent process. We offer design recommendations for AI-mediated tools that center disability perspectives and support interdependencies in job search. 
\end{abstract}

\ccsdesc[500]{Human-centered computing~Empirical studies in accessibility}
\ccsdesc[500]{Human-centered computing~Empirical studies in collaborative and social computing}
\keywords{blind and low-vision, job search, employment, job seeker, accessibility in employment, professional networking, social media, AI in employment}


\maketitle

\input{1_introduction}

\input{2_related_work}

\input{3_methodology}

\input{4_findings}

\input{5_discussion}
\input{6_conclusion}

\bibliographystyle{ACM-Reference-Format}
\bibliography{lit1}

\end{document}

%% file: 1_introduction.tex
\section{Introduction}
Blind and low-vision (BLV) individuals face persistent difficulties securing employment \cite{afb_visual_disability_employment}. Recent data indicate that approximately 55\% of BLV individuals in the United States are not employed \cite{afb_visual_disability_employment}. Among those actively seeking work, the unemployment rate is about 8\% \cite{nrtc_unemployment_visual_impairment}, roughly double that of people without disabilities \cite{bls_disability_unemployment2024}. A key contributor is that traditional off-line job-search techniques\footnote{In this paper, ``job search” refers to the full process: finding opportunities, applying, interviewing, and securing employment.} often fail to meet BLV job seekers’ needs \cite{grussenmeyer_2-evaluating_2017}. As a result, many BLV job seekers have turned to online job platforms and developed new strategies for navigating digital job search \cite{voykinska_9-how_2016, wu_11-visually_2014}. However, online platforms also pose accessibility barriers. One study found that 97\% of home pages violate the Web Content Accessibility Guidelines (WCAG) \cite{reuschel2023accessibility}, which could hinder BLV job seekers’ independent navigation of job sites and their ability to complete online applications. This may explain why only 28\% of BLV job seekers are able to complete job applications on their own \cite{lazar_1-investigating_2012}. Many BLV individuals continue to face challenges in securing employment, even as they engage with the online job ecosystem \cite{lazar_1-investigating_2012, grussenmeyer_2-evaluating_2017}.

Part of the problem is that the modern job search is also becoming increasingly complex as employers rely on AI-driven hiring systems, which can create new barriers for BLV individuals \cite{Nugent2021Recruitment, buyl2022tackling}. These AI systems often pre-screen applicants and decide who advances to interviews \cite{armstrong2025navigating}. Unfortunately, many of these algorithmic systems can disadvantage BLV candidates \cite{buyl2022tackling}, especially when their application portals are incompatible with screen readers or lack accessible labels on form elements \cite{glazko_identifying_2024, buyl2022tackling}. Recent work has also shown that AI-based résumé screeners can exhibit biases against disability-related achievements \cite{glazko_identifying_2024, buyl2022tackling}, potentially filtering out qualified BLV candidates before human evaluation. These preliminary findings are both alarming and important. However, while prior work has examined BLV individuals' experiences with online job platforms and accessibility barriers \cite{grussenmeyer2017evaluating, lazar_1-investigating_2012}, there remains limited empirical understanding of how they navigate the growing landscape of AI-mediated hiring. 
We know little about the specific challenges they face or the opportunities they identify when navigating AI-mediated hiring. Gaining deeper insight into these dynamics is key to avoid reinforcing systemic barriers, inform policy and practice, and to design accessible, worker-centered tools that can better support BLV job seekers. Our study is guided by the following research question:
\begin{itemize}
    \item \textbf{RQ:} How do BLV job seekers experience and navigate AI-mediated hiring systems?
\end{itemize}

To address this question, we conducted semi-structured interviews with 17 BLV job seekers and analyzed their experiences using the interdependence framework, which foregrounds how access emerges through relationships among people, technologies, and environments rather than individual effort \cite{bennett2018interdependence}. We chose this lens because AI hiring is not a simple interaction between one applicant and one tool \cite{armstrong2025navigating}. Outcomes can depend on how recruiters, platform policies, résumé parsers, screen readers, CAPTCHA and security mechanisms, and organizational timelines fit together \cite{li2021algorithmic}. An individual lens risks attributing breakdowns to a BLV applicant’s skills or effort, rather than to mismatches among how these different components operate together \cite{bennett2020care}. The interdependence framework instead, directs attention to these relationships and to the coordination work required for people to be able to access and participate in the hiring process. Note that by ``access,” we mean the practical ability of BLV individuals to participate in hiring, for example perceiving job content with a screen reader, navigating application forms that do not time out or block assistive technologies, submitting job materials that are faithfully interpreted by parsers, and receiving feedback they can read and act on. We believe this lens is necessary to understand BLV experiences within AI hiring systems because it can help capture where access succeeds or fails in the links among people, technologies, and environments, and it can help surface the collective strategies BLV job seekers use to make those links work.

Through our study we uncovered that BLV individuals reported that AI hiring systems often misrepresented their professional identities and produced dehumanizing, contextless interactions. In response, BLV job seekers engaged in strategic counter-navigation, deploying their own tools to bypass algorithmic filters and building peer networks to share AI literacy, and at times practiced strategic refusal, opting out of particular AI systems to reclaim agency. Contrary to portrayals of BLV job search as an individualistic effort focused on compliance, where job seekers are expected to strictly follow job ads, match keywords, and conform to whatever the employer requests, our findings show that job search in this context is an interdependent process. We conclude with design recommendations for AI job-search tools that center disability perspectives, scaffold interdependence, and intentionally enable strategic counter-navigation to level the playing field. This paper contributes to the field of human-computer interaction (HCI) in the following ways:

\begin{itemize}
    \item We present an empirical analysis of how BLV individuals navigate AI-driven hiring processes, showing that their job search in this context is not only an individual effort but also a collective activity that includes strategic counter-navigation.
    
    \item We highlight key limitations of AI hiring systems for BLV job seekers, including the ways in which these systems can misrepresent their professional identities.
    
    \item We propose design implications to improve inclusivity and usability in AI-mediated job systems. Specifically, we recommend creating tools that support collaboration in the job search process, as well as facilitate strategic counter-navigation, the deliberate use of alternative tools and collective strategies to navigate around algorithmic barriers, empowering BLV job seekers to shape fairer and more accessible job opportunities. 
\end{itemize}

By identifying current challenges and opportunities, as well as offering design recommendations, our research aims to guide the creation of more inclusive AI systems that support the employment needs of BLV individuals. This work aligns with broader goals in HCI to design equitable technologies that accommodate diverse user needs, particularly as AI has become increasingly central to hiring processes.

%% file: 2_related_work.tex
\section{Related Work}

Our work builds on research into accessibility in job search processes for people with disabilities (PWD), alongside studies on social media and AI's impact on job search and hiring.

\subsection{Job Search Accessibility}

The accessibility of online job platforms poses ongoing challenges for BLV job seekers, with studies highlighting persistent barriers such as incompatible screen readers, complex navigation, and lack of alternative text \cite{grussenmeyer_2-evaluating_2017, lazar_1-investigating_2012}. This leads to low application completion rates among BLV users, as many major job sites remain inaccessible despite established guidelines like the Web Content Accessibility Guidelines (WCAG) \cite{webaim2023, reuschel2023accessibility}. Although some tools, like CareerConnect® \cite{WOLFFE20051200} and ``Job Assist" \cite{brinkley2017case} have been developed to support BLV job seekers, they remain limited in scope and adoption \cite{WOLFFE20051200, brinkley2017case}. Our study expands on this previous research by examining in more depth how BLV job seekers interact with digital technologies during their job search. Specifically, we explore the strategies BLV individuals use to navigate digital platforms, including the creative workarounds they develop to overcome accessibility challenges, the methods they use to build professional networks, and approaches to personal branding as they strive to secure their desired jobs.

\subsubsection{Accessibility for Underserved Job Seekers}
Research on job search accessibility reveals that underserved groups, including people with disabilities, low-income job seekers, older adults, and racial minorities, often encounter structural inequalities, inaccessible digital tools, and lack of tailored support \cite{dillahunt_7-designing_2016, wheeler_3-navigating_2018, brewer_6-why_2016}. Studies suggest that while online platforms provide job listings, they fall short in guiding underserved users through effective job search strategies or accommodating specific needs \cite{marathe2025accessibility}, thereby perpetuating exclusion in the digital job search process \cite{harrington_9-working_2023}.

Our research adds to this conversation by focusing specifically on BLV job seekers and understanding in more depth their strategies to overcome barriers and advocate for equitable access in employment.

While much accessibility research focuses on developed nations \cite{sandnes2022there}, BVI job seekers in developing countries face 
additional barriers related to economic stratification, limited accessibility infrastructure, and societal biases \cite{kolgar2025insights, bhuiyan2025non}. Recent work examining BVI job seekers reveals how self-reflection becomes a critical yet challenging process, as individuals must navigate between high self-efficacy and practical skill gaps, often without adequate feedback mechanisms \cite{kolgar2025insights}. 
This work highlights how job intervention tools designed for Global North contexts may not address the unique needs of BVI individuals in linguistically diverse, resource-constrained environments \cite{dillahunt_7-designing_2016, cmar2021job, pal2012assistive}. Our study complements this research by examining how AI-mediated hiring, which is increasingly deployed globally, creates additional barriers that intersect with these existing challenges.

Recent work reveals that accessibility challenges continue even after employment begins in tech companies and identifies an "accessibility paradox," the inherent tension between organizations’ productivity- and profit-driven nature and their stated commitment to hiring and retaining disabled workers \cite{marathe2025accessibility}.

Our work examines the preceding stage: how BLV job seekers navigate AI-mediated hiring to reach employment in the first place. Together, these studies reveal a range of accessibility barriers across the employment lifecycle. Understanding both stages is essential for designing interventions that support BLV individuals throughout their careers.

\subsection{Social Media}

\subsubsection{Social Media Accessibility for BLV Users}
While social media offers opportunities for BLV users to connect and build networks, accessibility challenges persist. Studies indicate that many social media platforms lack consistent support for assistive technologies, which limits BLV users' ability to participate fully \cite{wentz_6-are_2011, whitney_5-am_2020}. Although some BLV users adopt creative workarounds, such as audio-based social media or assistance from sighted users, these adaptations highlight the ongoing need for more accessible digital spaces \cite{voykinska_9-how_2016, vashistha_10-social_2015}. Our study builds on this previous research by examining how BLV job seekers leverage social media platforms during their job search process.

\subsubsection{Social Media in Job Search and Branding}
Social media has transformed job search and recruitment processes, especially on platforms like LinkedIn, where personal branding and strategic engagement enhance visibility and networking opportunities for job seekers \cite{Basu2021, Johnson2021, mowbray_6-using_2021}. Younger generations increasingly view personal branding as essential for competitive job markets, though they often encounter challenges aligning digital presentation with professional norms \cite{Trang2023I}. Social media also enables "participatory branding," where individuals co-create their brands with audiences through interactive platforms, offering a more dynamic approach to professional identity \cite{meisner_participatory_2022, chen_tweet_2021}. Our study now offers valuable insights into how BLV job seekers adapt their strategies to overcome challenges related to visibility, establish meaningful connections, and potentially improve their marketability during their job search.

\subsection{AI Accessibility Across Domains: Implications for Job Search and Hiring}
AI has introduced efficiencies in recruitment, yet growing evidence suggests it may reinforce biases \cite{li2021algorithmic}, particularly against people with disabilities \cite{buyl2022tackling}. Studies show that AI-driven recruitment tools often fail to accommodate PWD’s unique experiences, misinterpreting employment gaps or alternative career paths as inefficiencies \cite{Nugent2021Recruitment, Amin2022A}. Recent research reveals that AI can devalue resumes with disability-related achievements \cite{buyl2022tackling}, reflecting societal biases within the training data \cite{10.1145/3630106.3658933}. Inclusive design practices that involve people with disabilities in AI development have been suggested to address these biases and improve AI's fairness and accountability \cite{Trewin2019Considerations}.

Across domains, research on BLV users’ interactions with AI tools reveals similar accessibility challenges. AI coding assistants can both support and exacerbate barriers for BLV developers, prompting strategies such as “AI timeouts” to manage cognitive overload \cite{flores2025impact}. Blind users approach GenAI with “everyday uncertainty,” requiring cross-checking across sources \cite{tang2025everyday} and extensive verification work when systems misfit with blind users’ practices \cite{alharbi2024misfitting}. Screen reader users likewise face increased time and effort with AI-powered productivity tools \cite{perera2025sky}. These findings indicate that AI accessibility challenges extend across domains. Our study contributes to this area by exploring the experiences of BLV users with AI-driven job search and hiring tools, and also understanding their perceptions of these technologies.

\subsection{From Independence to Interdependence in Job Search}
Traditional assistive technology research has emphasized independence, enabling individuals to complete tasks autonomously without assistance \cite{bennett2018interdependence}. While this focus emerged from important disability rights efforts against institutionalization, it may inadvertently obscure the collaborative nature of how access is actually created. \citet{bennett2018interdependence} proposes interdependence as a complementary framework that recognizes how all people rely on networks of support, tools, and relationships to accomplish goals \cite{bennett2020care}.
The interdependence framework reveals four key insights: (1) it focuses on relations between people, technologies, and environments rather than individual capabilities, (2) it recognizes multiple forms of assistance happening simultaneously, (3) it makes visible the contributions of people with disabilities in creating access, and (4) it challenges hierarchies that privilege certain forms of ability \cite{bennett2018interdependence}. These principles are particularly relevant for understanding digital job searching, which inherently involves managing relationships with recruiters, AI systems, professional networks, and online communities \cite{trimble2011role}.

For BLV job seekers, these interdependencies become more complex and visible. They must orchestrate additional relationships with screen readers, sighted allies who assist with visual elements, disability communities sharing workarounds, and AI systems that may not accommodate non-traditional career paths \cite{10.1145/3630106.3658933}. Rather than viewing these as dependencies, interdependence recognizes them as collaborative access-making. This lens reveals the often-invisible work BLV job seekers perform: building trusted assistance networks, developing collective strategies, and creating accessibility knowledge that benefits others \cite{wheeler_3-navigating_2018, herskovitz2023hacking}.

Similar patterns appear in workplace settings. \citet{shinohara2021burden} documented a "burden of survival" among BLV doctoral students, who expended unaccounted-for labor managing accommodations and assembling workarounds. Prior work also found that BLV professionals must navigate "sociomaterial configurations of access" where technical breakdowns require coordination with sighted colleagues \cite{das2019doesn}. Research on collaborative ideation extended this, showing that disabled professionals must navigate inaccessible technologies while also braving ableist collaboration norms \cite{das2024comes}. These findings suggest that interdependent labor persists beyond hiring into employment itself \cite{branham2015invisible}.

Interdependence also helps us understand how BLV job seekers navigate AI-mediated hiring. Beyond individual accessibility concerns, this framework reveals how they collectively learn AI patterns, share gaming strategies, and create alternative networks that bypass discriminatory systems. These practices represent interdependent workarounds rather than individual workarounds \cite{bennett2018interdependence, bennett2020care}. By adopting this lens, we can better understand how BLV job seekers don't just adapt to inaccessible systems but actively work together to create access in digital employment landscapes.

%% file: 3_methodology.tex
\section{Method}
Our objective was to study the challenges and strategies that blind and low-vision job seekers encounter when navigating AI-driven hiring systems and digital job search platforms. To study this, we conducted interviews that received approval from our university's Institutional Review Board (IRB).

\begin{table}[htbp]
   \small
   \centering
   \caption{Demographic Details of 17 Participants (7 Employed, 10 Unemployed). All names are pseudonyms.}
   \label{tab:participant2}

   \begin{tabularx}{\textwidth}{|X|X|X|X|X|}
       \hline
       \textbf{Name} & \textbf{Self-reported Visual Disability} & \textbf{Employment Status} & \textbf{Jobs Applied for in the Past Year} & \textbf{Social Media Usage Frequency for Job Purposes} \\
       \hline
        P1  & Blind & Employed in a full-time job & 0 & Weekly \\
       \hline
       P2 & Blind with light perception & Employed as a contractor & 0 & Weekly \\
       \hline
       P3 & Totally blind & Employed in a full-time job & 1-5 & Weekly \\
       \hline
       P4  & Blind & Employed in a full-time job & 0 & Never \\
       \hline
       P5 & Totally blind & Employed in a full-time job & 0 & Daily \\
       \hline
       P6  & Blind & Employed in a full-time job & 6-10 & Daily \\
       \hline
       P7  & Limited residual vision (LCA) & Employed in a full-time job & 6-10 & Monthly \\
       \hline
       P8  & Low vision and Stargardts & Unemployed and participating in a temporary work & 11-20 & Weekly \\
       \hline
       P9  & Retinal degeneration, glaucoma, low vision & Unemployed & 20+ & Daily \\
       \hline
       P10 & Light perception only (ONH) & Unemployed & 11-20 & Monthly \\
       \hline
       P11  & Low vision & Unemployed & 20+ & Daily \\
       \hline
       P12  & Fully blind & Unemployed & 6-10 & Weekly \\
       \hline
       P13  & Legally blind (glaucoma, RM) & Unemployed & 11-20 & Weekly \\
       \hline
       P14  & Blind with light perception & Unemployed & 1-5 & Weekly \\
       \hline
       P15 & Totally blind & Unemployed & 6-10 & Weekly \\
       \hline
       P16  & Totally blind & Unemployed & 11-20 & Weekly \\
       \hline
       P17 & Legally blind & Unemployed & 6-10 & Monthly \\
       \hline
   \end{tabularx}

   \smallskip
    \small
    \noindent \textbf{Notes:}\\
LCA = Leber congenital amaurosis; ONH = optic nerve hypoplasia; RM = retinopathy maturity\\[6pt]
    \textbf{Social Media Usage Summary:} Daily (4), Weekly (9), Monthly (3), Never (1)\\[6pt]
    \textbf{Job Applications Range:} None (4), 1-5 (2), 6-10 (5), 11-20 (4), More than 20 (2)
\end{table}

\subsection{Participants}
We conducted interviews with a total of 17 blind and low-vision individuals who were actively searching for jobs and had experience using social media platforms, such as LinkedIn, Facebook, and X (formerly Twitter), for job search purposes. Participants were recruited through local community partners via online mailing lists targeting screen reader users. Additionally, we utilized social media recruitment flyers and post-interview snowball sampling to reach a broader pool of participants. Our selection criteria for participation in the study was participants: (a) self-identified as blind or low-vision; (b) were 18 years of age or older and, (c) and were able to communicate in English. Additionally, they had actively searched for jobs using digital tools within the past year and used social media for job search purposes. Demographic details of the participants are summarized in Table~\ref{tab:participant2}.



\subsection{Procedure}
We conducted 17 one-on-one semi-structured interviews remotely over Zoom between June and September 2024. Each interview lasted an average of 63 minutes, ranging from 51 to 81 minutes. Participants were compensated at a rate of US\$30 per hour (prorated), with payment provided through either PayPal or Amazon gift cards. The interview protocol covered participants’ job search histories, the accessibility of digital job platforms and social media, the role of AI in the job search process, experiences with support resources, disclosure of disabilities, and the negotiation of accommodations with potential employers. Detailed notes were taken during each session, and all interviews were video-recorded and transcribed for analysis.




\subsection{Data Analysis}
We analyzed interview data using reflexive thematic analysis \cite{braun2021thematic}. One researcher open-coded all transcripts with attention to our research questions on BLV job seekers' experiences navigating AI-mediated hiring. Coauthors reviewed codes for accuracy and completeness during weekly meetings, where we discussed emerging patterns, refined the codebook, and resolved disagreements through consensus. Consistent with reflective thematic approach \cite{braun2021thematic}, we prioritized interpretive depth over inter-rater reliability. After open coding, we grouped codes into higher-level themes through axial coding, continuing until no substantially new insights emerged. Participant quotes are anonymized as "P" followed by a participant ID.




%% file: 4_findings.tex
\section{Findings}
Our analysis surfaced six key themes revealing how BLV job seekers navigate AI-mediated hiring: 

\subsection{Strategic Counter-Navigation Outsmarting Employer Filters with AI}
Our research uncovered that job-seekers who are visually impaired were strategic users of AI tools, which they leverage to navigate and counter perceived biases and structural barriers in AI driven hiring. Through our interviews, we found that participants developed deliberate tactics to bypass opaque algorithmic gatekeepers, e.g., automated resume screeners. They considered that these algorithmic gatekeepers might otherwise filter them out before a human ever reviews their application. In particular, participants described several strategies for optimizing their job applications. They used AI tools such as word cloud generators to analyze job descriptions and extract key terms, which they then incorporated into their resumes and cover letters to avoid being filtered out by automated systems. P8 illustrated this approach through their use of word cloud generators to systematically identify and incorporate high-frequency keywords from job descriptions: \emph{``...I can take a job description and paste it into a word cloud generator [a type of AI tool]. And this word cloud generator will tell me the frequency, what words appear, and how often... it's really easy for me to then look and see which words I should be putting in my cover letter [...] That has been a very [helpful AI tool]... Because reading the entire job description and trying to retain like ``Oh, I see this word a lot''... that can be kind of taxing. But if you just paste it into a word cloud generator, it gives you that list—boom—and it's in order of most used to least used [words]... I definitely need to use all those words cause they [the employer] clearly like them a lot.} Note that we interpret this behavior as ``strategic counter-navigation'' because we are observing job candidates re-purpose AI to reverse-engineer algorithmic screening logic and exploit its keyword heuristics with the purpose of potentially ensuring that their job applications pass opaque, non-accessible filters. By aligning language to what they believe AI algorithms' privilege, BLV job seekers are trying to circumvent structural algorithmic barriers.

Similarly, P7 explained that she deliberately customized her resume and cover letters to include specific keywords she believed were favored by employers’ AI screening systems: \emph{``I'm pretty sure there is AI used [in these hiring systems] [...] like if they're using it [AI] to help streamline their labor [labor associated to recruitment and hiring] [...] that's why I tailor...tend to tailor my cover letter and resume. Because I'm sure there's keyword filters, right? [keyword filters used for the hiring] Like [if you] don't have these requirements or meet the minimal qualifications or have these words—probably not getting looked at.''} We observe how this is also an act of strategic counter-navigation because P7 is consciously adapting her application materials to exploit the keyword filters of AI hiring systems, rather than presenting them in her own preferred style. By anticipating and shaping her documents around what the algorithm prioritizes, she aims to resist being excluded by automated gatekeepers.

Some participants took this ``strategic counter-navigation'' a step further. They reported utilizing generative AI to identify what hiring AI algorithms might prioritize and then they focused on creating job application materials that aligned with those inferred criteria. Essentially, they were using AI to outsmart AI screening.
P10 shared her approach to outsmart the algorithmic filters employers use to screen applicants:
\emph{“I've asked ChatGPT what keywords it would look for a certain job [...] I think companies specifically tell it [AI], like companies specifically use their [AI] settings to look for keywords [...] They could actually weed people out, like the best people for the job, just because they didn’t use the right buzzwords.”}. This meta-approach revealed participants' sophisticated understanding of how algorithmic bias could exclude qualified candidates based on language patterns rather than actual qualifications.

Overall, this theme helps to highlight how BLV individuals practiced strategic counter-navigation during their job search. They used AI to decode and exploit the very algorithmic filters that many times excluded them \cite{li2021algorithmic, buyl2022tackling}. Rather than passively accepting automated gate keeping, BLV job seekers treated keyword optimization and AI-assisted tailoring as tactical moves to push past opaque algorithmic screening. They did this with the purpose of securing a human who could review their job application materials. In doing so, they converted AI algorithms from barriers into systems they could outmaneuver.

\subsection{Dystopian Efficiency and The Dehumanizing Speed of Algorithmic Rejection}
Another theme that emerged from our interview data was about feeling dehumanized due to algorithmic rejections. BLV job seekers expressed a sense of powerlessness and frustration when confronted with what they perceived as instant and invisible rejections by AI-driven hiring systems. Unlike human-mediated processes, where context and individual circumstances might be considered for potentially hiring someone \cite{ling2025applicants, armstrong2025navigating}, our participants shared that the ``algorithmic gatekeepers'' rarely considered their unique backgrounds and instead operated with a binary efficiency that was dehumanizing as well as unfair. The speed of these rejections compound the emotional toll of job searching, created what participants described as a "dystopian" experience where their worth as potential employees was reduced to simple keyword matches and algorithmic calculations. P8 captured this sense of helplessness in the face of automated systems: \emph{``But it's harder for everybody [now that there is algorithmic hiring], because the employers can now just push a button and filter out all the candidates that they think they're not [qualified]. And you know you're gone. You're gone in a second, and it might be because you didn't use a certain word often enough. You know you might be perfectly qualified, but the bots [AI] don't know that."} 

This arbitrary nature of algorithmic rejection meant entire professional histories could be dismissed instantly based on missing keywords rather than actual qualifications. P1 also articulated how AI systems lack the human judgment that might recognize potential despite imperfect formatting or presentation: \emph{"If the resume is not properly given [properly formatted], if it's manually being, you know, done in a shortlist, for example. I think the interviewer [human interviewer] might think: ``okay, okay, this is fine. So still, we can consider this particular resume''. But AI will not listen like that. Right? AI will only go by certain rules, certain keywords. It [AI] will not have emotions."}

Through this, we observe how BLV job seekers started to articulate a clear distinction between human and AI evaluation. While a human recruiter might overlook minor formatting issues or recognize relevant experience despite imperfect keyword matches, our participants believed that AI systems operated without such flexibility. The emotional toll of this automated rejection became more evident as participants described how algorithmic screening undermined their self-worth and professional identity. P8 reflected on the ``soul-crushing'' nature of the modern job search: \emph{"It feels very dystopian to me, and that is significantly what I don't like about it [about AI hiring]. My worthiness as a human and as an employee, as a worker, is based on my ability to filter myself through an automated, a series of automated gateways. You know it has nothing to do with the job that I'm applying for."}

This disconnect between actual capabilities and rigid algorithmic criteria proved particularly painful for individuals with visual impairments, whose potential was misrepresented by systems that could not recognize their unique strengths.

We found that both “algorithmic dystopian efficiency” and accessibility barriers made participants question how employees who depended on these systems would treat them. They wondered what their own experiences might be like if they worked in companies that used such practices. For example, P6 reflected: 
\emph{"If I'm facing roadblocks at this, you know, at just these very beginning stages [during the job application phrase], it doesn't give me much hope... It just shows me that accessibility is not a priority for this organization [employer]...what other roadblocks would I face if I did accept the position?"}

\subsubsection{AI Rejection and Emerging Accessibility Barriers}
Our participants reported that AI-driven rejections created a new accessibility barrier: they could not reach a human reviewer who might consider their application in context. As P8 explained,: \emph{``I don’t even think a lot of people [BLV job applicants] have the opportunity to even reach a person [a human from the company seeking new employees]. So the inaccessibility is not even that a recruiter didn’t consider you. Tt’s that an algorithm [AI] filtered you out before you could even get to a human.''}

This barrier was compounded by worries about how AI screened applications. Participants believed that automated systems often evaluated visual features such as layout, formatting, or images during the first pass. For blind or low-vision job seekers, who may not perceive these cues, this meant being judged by standards they could not fully detect or understand. As a result, it was nearly impossible to adjust materials or take meaningful corrective action. For example, P7 noted:
\emph{``I know I’m probably missing some jobs that require visual materials or certain formats because the bots [AI] won’t pick me up I can’t even get to the point where I explain how I do things differently.”}

The experiences of our participants with dystopian efficiency helps to reveal a tension in contemporary hiring practices: While automated systems promise objectivity and efficiency, they often operate in ways that can feel arbitrary, exclusionary, and dehumanizing to job seekers, particularly those who already face additional barriers due to disability.

\subsection{Collaborative Learning and AI Hiring}
Peer-to-peer learning was also a central theme in our interviews. BLV job seekers built informal networks to share practical tips about AI hiring systems: which tools helped them navigate algorithmic screening, which settings worked, and how to avoid common pitfalls in algorithmic hiring. Instead of formal courses or institutional programs, they learned from friends, peers, and community contacts who were also experimenting with these technologies. As P13 noted:
\emph{``I would say the most [useful resource for navigating AI hiring] is probably people, either my friends or people I just know socially or from my student community. I think I found that more useful than the vocational or the career office [...] they [her network] have similar insights to current trends [trends in AI hiring] in terms of issues and benefits that I would need to know. So, they’ll [her network] give me information, or like a piece of advice... [that’s] more helpful...''}

This peer-to-peer knowledge transfer appears to represent a grassroots form of digital empowerment, where BLV individuals collectively navigate the evolving landscape of AI-enhanced job search. The advantage of these networks is that participants can gain insights from people with disabilities who are going through similar circumstances and learn creative ways of navigating the evolving technological landscape, even when they are not technology experts themselves. As P8 emphasized: \emph{"My community [BLV job seekers in the AI hiring ecosystem] is so important to me. I wouldn't be the person that I am today without having developed meaningful connections and networks within my blind tribe [...] That's where I've gotten mentoring. It's where I learn things all the time from other blind people's experiences [...] I'm kind of learning from my friends [learning about AI from her friends], like my friends, are doing a lot of creative stuff with it [with AI]..."}. 

These peer learning networks provided participants with technical knowledge, emotional encouragement, and professional guidance throughout their job searches. In particular, participants described using these networks not only to learn new skills but also to discover promising job opportunities. Because of this multi-purpose value, participants viewed their peer networks as one of their most effective resource for the job search process. For instance, P16 shared that she preferred relying on peer networks over more formal job search channels:
\emph{"I guess getting recommendations from friends [is my preferred method for learning about job opportunities]. Around where I live a lot of times I found out that if you know somebody who knows somebody, it can help you get in the door [and get the job]."}. We note that this informal learning network likely made the job search process easier by helping participants bypass application systems that were potentially inaccessible or filtered by AI, which may have been less accommodating to BLV applicants.

Participants also articulated how observing the career trajectories of people in their network provided inspiration and practical learning opportunities. As P5 reflected: \emph{"I'm connected to my friends, people I've known for a long time I'm connected to many people who are doing the same kinds of things that I'm doing [BLV individuals working in their same field] seeing the kind of path that she [woman in her network] takes and how she got there, and realizing that I'm kind of on the same sort of path, it keeps me inspired, and it keeps me believing and learning, because that's one of the most important things you can do."}. This observation-based learning extended to career strategy and professional development, creating a comprehensive support system.

On the other hand, it is important to note that while our participants emphasized that peer knowledge sharing was useful for learning new skills and techniques, they also recognized that limited networks could create disparities in access to information. For example,  P15 had the following experience: \emph{"I've not found [anyone who talks about job searching on LinkedIn], even in my circle of friends who are blind. That's not been something that I have had a lot of experience with on my own, nor have I had peers who talk a lot about LinkedIn."} As a result, while peer networks provided valuable support for those with access, gaps in these informal knowledge-sharing networks could leave some job seekers without critical information about AI tools and digital job search strategies.

\subsection{Misrepresentation in AI-Generated Job Search Support}
AI-generated content often misrepresented participants’ professional identities, forcing them to spend extra time correcting how the tools portrayed them and making the job search harder. Despite promises of personalization, participants felt that AI tools for CVs, cover letters, and bios frequently produced generic or misleading results that diminished, rather than highlighted, their skills and experiences. This was especially problematic for BLV job seekers, compounding the challenges of presenting themselves effectively in digital hiring systems. For example, P10 described how LinkedIn’s AI bio-writing feature failed to capture their background:
\emph{"I've tried that feature [LinkedIn's AI bio-writing]. But what it [AI] writes about me isn't very good. It might focus way too much on one detail, or it'll [the AI will] give like general things about me. It [The AI] can basically kind of look like you didn't do that much. It [The AI] keeps writing that I can code in programming languages that I can't code in like Java. I was trying to focus on how I was able to gather and process data. It [The AI] keeps saying that I'm good at psychology and mental health when that's not what I was trying to say..."}

In our interviews, we also observed that this AI misrepresentation problem extended beyond individual profiles to job matching algorithms. P16 described receiving irrelevant job recommendations, likely because the AI had misrepresented their skillset: \emph{"Sometimes they'll [the AI job recommendation system will] put social work jobs up just because I have a bachelor's in social work [...] or sometimes they'll [the AI job recommendation system will] have like nursing jobs. And I'm like, I'm not a nurse [...] I guess it's just an AI thing trying to go by whatever it is I've put up there [on their digital biography], and I guess it [AI] makes mistakes...".} This algorithmic mismatch can waste job seekers' time and can reflect a failure of the AI system to understand career trajectories and professional identity beyond keyword matching.

The misrepresentation by AI algorithms was also evident in the AI's failure to understand that certain jobs might not be suitable recommendations for a job seeker with visual impairments, particularly when those positions appeared to have accessibility issues that would make job performance difficult.  As P10 noted: \emph{"LinkedIn is basically the only thing I can use at this point [the only tool they use for job search], because Indeed [another online platform for job search] is not that accessible. It [Indeed's AI job recommendation system] keeps on giving me jobs that wouldn't be accessible..."}. The irony of an AI system recommending inaccessible jobs to a job seeker with visual impairments can help to highlight how these AI systems fail to represent essential context about users' needs and constraints.

P10 further theorized about why AI might fail at correctly representing the needs and characteristics of job seekers who are visually impaired:
\emph{"The way it [AI] thinks is, it [AI] just gets patterns from like, it's [AI has] gotten all the data from the Internet, and it [AI] gets patterns. And it [AI] can't like understand nuances. So basically, it's [AI is] like, Oh, well, you know, this [skill on your bio] was mental health related. So that must mean you are good at psychology."}

BLV job seekers' insight captures an important limitation of AI tools in hiring systems. Because they operate on pattern matching, AI systems usually cannot grasp the contextual nuances that define the professional identities of individuals. Participants' experiences with misrepresentation failures also reveal a gap between the promise and reality of AI-powered job search tools. Rather than providing tailored, intelligent assistance, these systems appear to be producing generic, inaccurate representations that could potentially harm job seekers' prospects. For job seekers with visual impairments, who already face additional barriers in conveying their qualifications and navigating inaccessible systems, these AI failures can represent yet another obstacle in an already challenging job search process.

\subsection{Strategic Refusal: Rethinking Agency in the Age of AI Tools}

While AI-powered job search tools have become increasingly prevalent, offering features like resume optimization, application automation, and job matching algorithms \cite{li2021algorithmic}, our interview study uncovered that not all job seekers embrace these technologies equally. For BLV individuals, the decision to use or avoid AI tools in their job search involves complex considerations beyond mere accessibility. This theme examines a notable pattern we observed among visually impaired job seekers: the deliberate choice to forgo AI assistance in their job search processes. This conscious non-adoption of AI appeared to emerge as a strategic response to the broader ecosystem of automated hiring practices, where AI-driven screening systems could perpetuate or amplify discrimination against candidates with disabilities. As P9 explained
\emph{“I choose to not use AI technology when I'm doing my job [job search process], because I would rather them [the employer] not use AI technology to screen me either [...] I hope that I can help set the standard for not using AI with job searches. [...] if they [the employer] have a real person going through the applications... they [the employer] might go ahead and call me for the interview regardless. But if they [the employer] have AI checking everything and screening out all the people who say no, they don't drive... Then I might be shoved into that pool of non interesting applicants. Simply because AI checked the box.”}. 

Participants who avoided AI tools succeeded by using traditional job search methods, though they recognized AI's potential effectiveness in other contexts. As P3 noted: \emph{"[I] did an informational interview [...], and then did one more [interview], and I got it so [I got the job]. No, I did not use AI tools [for the job interviews]. I did just have help from a family member to make a resume. But, I did use an AI once, and I was astounded at how effective it [AI] was. But other than that, I can't really say I've used AI for job stuff."}

This contrast between acknowledging AI's capabilities and choosing human-centered approaches can showcase that non-adoption is likely steming from deliberate choice rather than lack of awareness. 

Others felt that AI tools were insufficiently developed for meaningful integration into their job search process, leading them to avoid using them. For example, P15 expressed skepticism about the current usefulness of AI for job seeking, which is why they chose to limit their use of such tools:
\emph{"I don't use any AI... AI is not new, but still in its infancy. And so it's [AI has] not made its way into my job search space yet. I've not found that it's [AI has] been helpful. [...] There are some things, like algorithms [...] that you don't know of [don't know that AI algorithms were already present], that have been AI all along [features in a job search interface that has actually always involved AI algorithms], that match you with certain jobs and because of certain criteria [...] those weren't things that I initiated [he did not actively search for AI tools]. Just that's just the way that the tool is built [the tool has AI already integrated into it]"}. 

This perspective highlights how participants distinguish between AI they actively choose to use versus embedded algorithmic systems they cannot avoid. As a result, participants navigate a complex landscape where conscious non-adoption represents one form of agency in an increasingly automated hiring ecosystem.

\subsection{Accessibility by Chance, The Patchwork Nature of AI Tool Usability}
Through our interview, we also captured the unpredictable landscape of AI tool accessibility that visually impaired job seekers must navigate, and their reliance on assistive technologies to bridge these gaps. Participants described a reality where accessibility in the job search process appears more accidental than intentional, forcing them to develop complex workarounds using screen readers, voice assistants, and AI-powered visual description tools.

For example, P12 highlighted the inconsistency across AI tools:
\emph{"I've had some... AI apps that aren't accessible. Luckily, the ones I've used recently like Perplexity is accessible. ChatGPT is accessible. Copilot is accessible. I think there's others out there that... may not be."}
The word "luckily" reveals how accessibility is not an expected standard but a fortunate discovery. 

Even when AI tools were technically accessible, participants encountered compatibility issues with their assistive technologies. P10, who uses a NonVisual Desktop Access (NVDA) screen reader, which  is a free, open-source screen reader for Windows that allows BLV individuals to use computers through synthesized speech or braille displays, noted that while ChatGPT had become accessible, she remained cautious: \emph{"ChatGPT used to have accessibility problems, but... those have gotten fixed for now. I don't know if they're going to break accessibility again in the future."}

This uncertainty forced participants to remain vigilant about potential accessibility breakdowns that could disrupt their job search process.

The complexity of navigating AI accessibility in job search was compounded by the variety of assistive technologies participants needed to use across different devices and contexts. P7 described her multi-layered approach: \emph{"I use VoiceOver to interact with that content [job ads on  her iPhone]. When it comes to actually applying for the positions... I will go on my Windows computer/laptop. And I use JAWS [a type of screen reader], sometimes NVDA [another type of screen reader], but mostly JAWS screen reading software to then complete the application.". We note how this multi-layered approach required constant switching between technologies depending on the task and platform.}

When AI tools proved inaccessible through traditional screen readers, participants turned to AI-powered visual assistance tools as workarounds. P10 explained using ``Be My AI'' and ``Aira Access'' to describe inaccessible content: \emph{"I will share the photo to my photos they have on Be My Eyes. They have one [one feature] named Be My AI, and Aira got AI as well named Aira Access... with Aira Access, you can actually call a live human to see if what the AI said was correct."}

Participants also described how they used AI to help them navigate job search platforms that were not as accessible or they had confusing layouts. P14 for example, described using Be My AI to navigate inaccessible forms: \emph{"I just moved my finger on my computer's mouse, and then take a picture of the screen and let AI describe where the cursor is... and then it tells me around how far to move my finger, or what direction."}

We note that participants had to engage in a kind of patchwork work to navigate an AI-driven job search process that was often inaccessible. Because of this fragmented landscape, they also had to develop their own verification strategies to test whether a tool or platform was reliable before deciding to commit to it. As P12 advised: \emph{"You always want to make sure that you do extra research on the side off of the AI apps just to make sure... you're getting something that's actually going to be resourceful."}

We observe how visually impaired job seekers were investing significant time and effort navigating between multiple assistive technologies and AI tools, while their sighted competitors likely could readily access AI-powered job search advantages directly, creating a form of technological inequality in the hiring process.

%% file: 5_discussion.tex
\section{Discussion}

Through our study, we are uncovering that job search for Blind and Low Vision (BLV) individuals within AI hiring systems is far from the isolated and linear process that prior accessibility research has typically examined \cite{grussenmeyer2017evaluating, reuschel2023accessibility}. Instead, our findings reveal multiple dimensions of interdependence at work. Consistent with \citet{bennett2018interdependence}'s framework, which emphasizes relations between people, technologies, and environments rather than individual capabilities, we found that job search for BLV individuals is a collaborative process involving constant interaction with networks of peers with shared experiences, challenges, and strategies. BLV job seekers rely on mutual support and shared knowledge on how to navigate complex hiring infrastructures, often exchanging tips on accessible job boards, compatible assistive technologies, and the quirks of AI systems. Job search also requires engagement with multiple platforms, assistive technologies, and AI tools simultaneously. BLV individuals often switch between different devices, browsers, and AI assistants to complete even simple application tasks\cite{ubiquitousmultiplescreenreaders, flores2025impact} . These practices make visible the contributions of people with disabilities in creating access: the collective knowledge-sharing and workaround development we observed represent active access-making labor rather than passive adaptation \cite{bennett2020care}. This interdependence of systems and people turns job seeking into an ecosystem of collective adaptation and strategic navigation, not an individual effort based solely on compliance.

The grassroots strategies participants described are not failures to fix but models to build upon. They demonstrate how marginalized communities like BLV create their own interdependent infrastructures when formal systems exclude them \cite{dillahunt_7-designing_2016, dillahunt_5-examining_2021}. As AI becomes increasingly central to employment access \cite{armstrong2025navigating}, supporting these forms of interdependence becomes crucial for equity. Finally, the interdependence we observed extends beyond job search into employment itself. \citet{marathe2025accessibility} describes the "accessibility paradox," where organizations publicly commit to accessibility while systemic barriers persist. Our findings reveal that this paradox begins even before hiring. Addressing it requires moving beyond compliance-based accessibility toward understanding interdependence as fundamental to how all people navigate technological systems.

\subsection{Recognizing Interdependence as Infrastructure in AI Mediated Job Hiring}
Our findings reveal that interdependence functions as an infrastructure for employment access. BLV job seekers depend on relational networks and coordinated technologies not only when requesting accommodations, but throughout the entire job search process: from learning about a job opportunity, to completing applications, to bypassing algorithmic filters, and preparing for interviews. However, current AI hiring systems create opaque, inaccessible algorithmic systems that systematically disadvantage BLV job seekers \cite{buyl2022tackling}. 

While peer networks benefit all job seekers \cite{obukhova2013job}, the nature of interdependence differs fundamentally for BLV job seekers. For sighted job seekers, peer support enhances efficiency but is not prerequisite for basic participation. For BLV job seekers, interdependence is often mandatory infrastructure rather than optional enhancement. \citet{shinohara2021burden} documented how BLV doctoral students face a "burden of survival" through unaccounted-for labor to bridge inaccessibility. \citet{das2019doesn} found that blind professionals must navigate "sociomaterial configurations of access" where technical breakdowns require coordination with sighted colleagues simply to participate in routine activities.

AI hiring system magnifies these disparities. Coordination costs are exponentially higher as BLV job seekers may have to simultaneously orchestrate screen readers, multiple AI tools, and peer knowledge about accessible platforms before submitting a single application \cite{perera2025sky}. The stakes of failure are also asymmetric: when a sighted job seeker's network fails, they can still independently complete applications; when a BLV job seeker's assistive technology ecosystem fails, they may be entirely locked out. Additionally, AI hiring introduces new layers of opacity as BLV job seekers cannot visually inspect whether their resume renders correctly or verify whether AI-generated content accurately represents them \cite{alharbi2024misfitting}. This dual burden of managing both technical inaccessibility and ableist professional expectations \cite{das2024comes} represents a  different form of interdependence than what sighted job seekers experience.

The dystopian efficiency experienced by participants represents a critical infrastructure failure. Unlike human-mediated processes where context and nuance might be considered, algorithmic systems operate with binary efficiency. While sighted candidates might quickly adjust applications based on visual cues, BLV job seekers face additional layers of technological mediation that make real-time adaptation nearly impossible. The systematic misrepresentation of professional identities demonstrates another infrastructure failure. AI-generated professional summaries consistently miscoded skills, invented non-existent competencies, and failed to capture the nuanced career trajectories that BLV professionals navigate. 

While prior work shows that when workplace technologies are inaccessible, blind employees are forced to ``create their own accommodations”, extra efforts that constitute invisible labor to make systems workable \cite{branham2015invisible, xiao2024systematic}. This invisible labor (e.g. assembling workarounds, seeking help, using personal time to adapt resumes for algorithms) is a direct consequence of ``hostile” or opaque systems that assume an able-bodied, independent user \cite{li2021algorithmic, rocha2025awareness, mcdonnall2025one}. In our study, participants similarly described strategic assemblages of screen-readers, AI tools, and community advice for AI literacy as essential infrastructure to navigate AI hiring. 

Our focus on interdependence complements recent work examining self-reflection strategies among BLV job seekers \cite{kolgar2025insights}, which found that job seekers view reflection as an inherently social process requiring peer feedback and mentor guidance. While that work identified challenges in obtaining adequate feedback from human interviewers and mentors, our findings reveal an additional layer of AI-mediated systems that further diminish opportunities for human interaction, which are essential for productive self-reflection. The ``dystopian efficiency” described by participants in our study, where algorithmic rejection occurs instantly and without human review \cite{buyl2022tackling}, effectively removes the feedback loop that is crucial for BLV job seekers’ skill development and career decision-making \cite{kolgar2025insights}.

Policy frameworks must evolve to recognize the interdependent nature of AI-mediated hiring and the critical role of community networks \cite{wheeler_3-navigating_2018}. Current regulations for accessibility focus on individual accommodation and technical compliance \cite{ADA_Web_Accessibility_Rule_2024}, missing how systemic barriers emerge from forced interdependence with hostile, opaque algorithmic systems.
Regulations should require transparency about AI screening systems and their interdependencies with other technologies. The strategic use of generative AI tools by BLV job seekers in our study to understand opaque interdependent relationships between their application and algorithmic evaluation demonstrates this need. Policy could mandate that employers disclose not just that AI is used, but how it evaluates candidates and what interdependencies it assumes. This transparency would allow job seekers and their networks to make informed decisions about how to engage with these systems.
Community networks that participants relied upon for AI literacy and job search strategies represent critical infrastructure requiring policy support. Rather than funding only formal training programs that participants found less useful, policy should recognize and resource peer networks as essential support systems. This includes funding for peer mentorship programs, community-based AI literacy initiatives, and platforms for knowledge sharing among disabled job seekers.
Finally, policy must recognize the labor involved in maintaining interdependent relationships with inaccessible systems. Participants spent significant time assembling technological workarounds, learning from peers, and strategically adapting to algorithmic requirements. While HCI researchers have critically examined invisible labor among gig workers \cite{2024gigsense, toxtli2021quantifying}, for BLV job seekers, this form of interdependent labor remains largely invisible and uncompensated. Policy frameworks could compensate or mitigate this labor, for example by allowing extended time for AI-based assessments, or by holding companies accountable for the cost their inaccessible systems impose on applicants \cite{branham2015invisible, lazar2015ensuring}. In summary, a shift from viewing access solely as an individual responsibility to viewing interdependence as shared infrastructure can drive more equitable AI-mediated hiring practices.

\subsection{Exercising Agency in Technology Refusal}
Participants' conscious non-adoption of AI tools represents a strategic exercise of agency through selective interdependence rather than withdrawal from technology. Several participants articulated a philosophy of reciprocal refusal, choosing not to use AI in their job search because they preferred employers not to use AI to screen them. This orientation aligns with HCI scholarship that frames refusal and non-use not as failure or inability \cite{satchell2009beyond}, but as an intentional political and ethical stance toward technology \cite{cha2025understanding, costanza2020design, fuchsberger2014human}. In this sense, agency is not about achieving independence from technology, but about deliberately deciding which technological relationships to enter and which to refuse \cite{bennett2018interdependence}.

This selective engagement reveals sophisticated decision-making about technological relationships \cite{baumer2025exploring, cha2025understanding}. Participants recognized that engaging with AI hiring systems meant accepting forced interdependence with tools that might discriminate against them. Prior HCI work on algorithmic hiring similarly shows that job seekers are acutely aware of automated screening systems and actively adapt their behaviors to avoid being filtered out, for example by tailoring resumes to inferred keyword-based criteria or seeking alternative pathways into organizations \cite{lashkari2023finding, li2021algorithmic}.

Rather than passive acceptance, participants made calculated choices about which systems deserved their participation. Some found success through informational interviews and personal references, demonstrating how human interdependencies can circumvent algorithmic gatekeepers \cite{deck2024implications, tilmes2022disability}. Even those who avoided AI in job applications maintained interdependencies with their community for knowledge and support, learning from peers who were experimenting creatively with these technologies. This challenges narratives that frame technology refusal as inability or ignorance, revealing it instead as strategic navigation of power relationships \cite{costanza2020design}.

While strategic resume tailoring and professional networking are common across all job seekers, our findings suggest that BLV job seekers' technology refusal and counter-navigation practices are shaped by disability-specific conditions. The stakes of refusal differ: sighted job seekers who choose not to use AI tools retain full access to application systems, whereas BLV job seekers' "refusal" is often entangled with exclusion, as they may reject tools that were never fully accessible to them \cite{shinohara2021burden, das2019doesn}. Similarly, strategic counter-navigation also requires recursive technological dependencies as sighted job seekers can visually scan job postings to identify keywords, BLV participants needed to deploy AI tools to decode other AI systems, creating layers of mediation that amplify both labor and risk of errors \cite{flores2025impact, perera2025sky, ubiquitousmultiplescreenreaders}. The philosophy of "strategic refusal" articulated by our participants reflects awareness of AI systems' documented biases against disability \cite{buyl2022tackling, glazko_identifying_2024}. Additionally, BLV peer networks circulate accessibility-specific knowledge, such as which platforms are screen-reader compatible and which AI tools misrepresent professional identities \cite{perera2025sky, 10.1145/3663548.3675631, zhao2025use}, knowledge that sighted job seekers can largely take for granted.

The peer networks described in our findings represent alternative interdependent infrastructures that operate outside formal employment systems \cite{bennett2020care}. Participants emphasized that personal connections often provided the most effective pathway to employment, with informal recommendations helping candidates bypass automated screening \cite{lashkari2023finding}.

BLV job seekers' strategic use of AI to navigate AI screening systems demonstrates what we conceptualize as strategic counter-navigation. Rather than individual adaptation, participants engaged in collective navigation strategies that depended on both technological tools and community knowledge. Through Bennett et al.'s \cite{bennett2018interdependence} lens, we see this not as individual job seekers gaming the system, but as networks of interdependent actors sharing tools, strategies, and knowledge to navigate algorithmic gatekeeping.

This pattern aligns with findings from \citet{gamage2023blind}, who show that BLV users navigate assistive AI through selective engagement. Their participants often rejected technologies that failed to reflect their everyday priorities, choosing instead those that supported conversational and context-aware interaction. Similarly, our participants' selective use of AI hiring tools reveals a deliberate negotiation of interdependence rather than technological resistance. This interdependence extended beyond human networks to include the AI tools themselves, as participants used generative AI to understand what keywords hiring algorithms might prioritize, creating recursive loops of AI interdependence \cite{li2021algorithmic, mayworm2024content}.

\subsection{Design Implication}
Based on our findings, we propose design implications for a hiring system that moves beyond ``accessible AI" toward ``interdependent AI"- hiring systems that acknowledge and support the collaborative nature of job search and  access.

\textbf{Design for Transparent Interdependencies:} Hiring Systems should make their technical and social dependencies visible rather than hiding them behind claims of automation and efficiency. This means explicitly showing what file formats, assistive technologies, and human interventions the system expects or supports. Concretely, systems should:

\begin{itemize}
    \item Provide an ''interdependence maps'' that indicates where AI screening occurs, what stage the application is currently in, and what input formats and assistive technologies are supported, allowing candidates to understand where in the interdependent chain breakdowns might occur.
    \item Offer an application ''preflight'' mode that shows how resumes and forms are parsed before submission (e.g., extracted fields, detected keywords, missing sections, formatting issues) so applicants can correct problems proactively rather than submitting into an opaque filter.

    \item Ensure that system feedback is actionable and perceivable (e.g., readable error messages, explicit time limits, accessible confirmation that materials were received and processed). Provide human recourse pathways that are reachable without excessive friction, especially when automation rejects or blocks an applicant. Human recourse should not be an aspirational back-end process; it should be an accessible, user-visible optio
\end{itemize}

\textbf{Recognize Strategic Counter-Navigation and Refusal as Accessibility Signals:} Employers and platforms should treat strategic workarounds as signs of system accessibility failure, not problems to eliminate. For example:
 \begin{itemize}
     \item Drop-offs at specific form steps, repeated resubmissions, or high rates of opt-outs from AI-mediated assessments can be interpreted as accessibility telemetry. 
     \item Tracking when candidates use external AI tools to navigate application systems indicates that the system's own interdependencies with assistive technologies have failed.
 \end{itemize}

\textbf{Support reliable self-representation in AI-assisted job-search tools:}  Participants described AI tools that miscoded skills or produced generic portrayals. Platforms offering AI writing or recommendation features should incorporate safeguards that help BLV job seekers verify and correct outputs:
\begin{itemize}
    \item Interfaces that support structured review (e.g., claim-by-claim verification prompts, highlighting inferred skills, and easy correction workflows).
    \item Controls that allow users to constrain generation to resume-grounded evidence rather than speculative additions, and transparency about what sources were used. 
\end{itemize}



These principles recognize that designing for interdependence is not about adding features but fundamentally reconceptualizing how hiring systems relate to the communities they serve. By acknowledging that job seeking happens through interdependent relationships with technologies, communities, and institutions, we can create systems that support rather than undermine the collective strategies marginalized communities develop for job access.

\subsection{Limitations and Future Work}
While our study provides valuable insights into BLV job seekers' experiences navigating AI-mediated hiring, several limitations should be acknowledged. Our participant pool primarily included BLV individuals who are relatively comfortable with digital technologies, potentially missing perspectives of those with lower digital literacy whose experiences with AI hiring may differ significantly. Our interviews also captured a snapshot during rapid AI advancement and shifting remote work policies. Additionally, our focus on job seekers' perspectives means we lack direct insight into how employers perceive and respond to the interdependent strategies we identified.

These limitations suggest several directions for future research. Studies should examine how BLV individuals without strong community networks or technological expertise navigate AI-mediated hiring. Longitudinal research could track how interdependent navigation strategies evolve as AI systems change. Future work could also examine whether the counter-navigation strategies we document are unique to BLV job seekers or reflect broader patterns of algorithmic resistance among all job seekers. Research should investigate the employer side, including how organizations perceive applications showing evidence of AI assistance and whether hiring managers recognize the additional labor required for BLV candidates.

%% file: 6_conclusion.tex
\section{Conclusion}
Our study reveals the challenges BLV job seekers face navigating AI-mediated hiring: while AI systems promise efficient job matching, they create problematic interdependencies through instant algorithmic rejection, systematic misrepresentation of professional identities, and patchwork accessibility that demands exhausting technological choreography. BLV job seekers navigate these challenges through strategic workarounds and peer knowledge networks, but remain disadvantaged by opaque screening systems and inaccessible infrastructure. Using Bennett's interdependence framework, we identified critical gaps in current AI hiring systems, particularly in transparency, temporal justice, and recognition of non-standard career paths. Our design recommendations address these gaps by centering BLV job seekers' own experiences and supporting the interdependent strategies they have developed. These recommendations call for AI hiring systems that make dependencies visible, leverage community knowledge, and preserve human review pathways. Our findings challenge the framing of accessibility as individual accommodation in AI hiring system, revealing instead how BLV job seekers necessarily rely on complex interdependencies with technologies, communities, and workarounds simply to participate in the job market.